\documentclass[aps,pra,twocolumn,showpacs,superscriptaddress]{revtex4-1}  
\usepackage{graphicx}  
\usepackage{dcolumn}   
\usepackage{bm}        
\usepackage{amssymb}   
\usepackage{amsmath}   
\usepackage{epstopdf}   
\begin{document}
\title{Generation of Pulse Sequences with Periodic Spectral Phase Masks}
\author{Itan Barmes}
\affiliation{Department of Physics and Astronomy, LaserLaB, VU University, de Boelelaan 1081, 1081 HV Amsterdam, The Netherlands}
\author{Axel Ruehl}
\affiliation{Deutsches Elektronen Synchrotron (DESY), Notkestrasse 85, 22607 Hamburg, Germany}
\begin{abstract}
We present a method for the generation of complex pulse sequences by using periodic spectral phase modulation. An analytical expression is derived for the temporal profile of such pulse sequences, which relates the temporal amplitudes and phases of the subpulses to the Fourier components of the periodic phase mask. We consider two families of periodic phase masks, namely, when a single period is an odd or even function, and discuss the differences between the resulting pulse sequences. The ability to generate pulse sequences with more degrees of freedom provides an alternative parametrization for investigating quantum coherent control landscapes. This is illustrated in the context of dark pulses in nonresonant two-photon absorption.
\end{abstract}

\maketitle

The generation and manipulation of ultrashort (femtosecond) laser pulses has opened new possibilities in studying and controlling light-matter interaction. Such pulses provide excellent temporal resolution that allows to investigate the dynamics of molecular processes with high temporal resolution~\cite{zewail2000}. Simultaneously, due to the Fourier relation, such pulses exhibit broad spectral coverage. This enables the simultaneous excitation of a quantum system via different pathways. In the field of coherent quantum control~\cite{Tannor1985,Brumer1992,Warren1993} shaping of femtosecond pulses is employed to steer a system to a desired final state by means of constructive and destructive quantum interferences between pathways leading to the same final state. Pulse shaping techniques are mainly based on a frequency domain manipulation using a spatial light modulator placed in the Fourier plane of a zero dispersion 4f grating-based configuration~\cite{Weiner2000}. This provides programmable spectral phase masks~\cite{weiner1990b} that change the temporal properties of the pulse without affecting the total energy. Similar schemes also allow shaping of the amplitude and polarization of the driving field~\cite{Brixner2001,Brixner2004}.

An important step in designing any coherent control experiment is to determine appropriate control parameters~\cite{Langhojer2005}. Searching through all possible spectral masks is not only experimentally impractical but also difficult to interpret physically. Parametrization of the spectral mask according a physically relevant function basis can reduce the number of dimensions of the experimental search and assist in understanding the underlying control mechanism. Even though the manipulation is conducted in the frequency domain, understanding the temporal waveform provides valuable information about the dynamics of the system~\cite{Trallero-Herrero2006}. For example, allowing a single parameter to control the chirp rate (quadratic phase) has been shown to either strongly enhance or completely eliminate the two-photon absorption (TPA) signal. This behavior was explained by the dependence of the time-dependent instantaneous frequency of chirped pulses~\cite{broers1992a,chatel2003}. More control parameters allow investigation of control landscapes with more general properties of the control mechanisms. Many of the experimental realizations of control landscapes were demonstrated with parametrization of chirp orders~\cite{Roslund2006,Walle2010}.

An alternative parametrization is a sinusoidal spectral phase in the form $\varphi(\omega)=A\sin[\omega T+\phi]$, where the control parameters $A$, T and $\phi$ determine the modulation amplitude, period and phase, respectively. It has been previously shown~\cite{wollenhaupt2005} that such a spectral phase mask results in a sequence of equally spaced subpulses, where the shape of the envelope of each subpulse is the same as the unmodulated pulse. This parametrization was employed to various systems from controlling biological systems~\cite{Herek2002}, selective excitation of molecular vibrational levels~\cite{Dudovich2002} to the spatial control of atomic excitation~\cite{Barmes2012}.

In this paper we generalize the concept of a single sinusoidal spectral phase modulation to arbitrary periodic phase masks. We derive an analytical expression for the temporal amplitude and phase profiles of the resulting pulse sequences. The derivation is based on the Fourier expansion of periodic functions where a single period is an even or an odd function. These two families of phase masks present unique properties compared to the case of a single sinusoidal phase mask. These properties can be exploited for enhanced controllability and provide an alternative parameterization for investigating control landscapes.

For sake of comparison we briefly summarize the properties of a pulse sequence when applying a single sinusoidal phase mask (for a detailed derivation see Ref~\cite{wollenhaupt2005}). In the time domain this corresponds to a sequence of subpulses separated by time T. The temporal amplitude of the n$^{th}$ subpulse is $|J_n(A)|$, where $J_n(A)$ is the Bessel function of the first kind of order n. The temporal phase displays different patterns for even and odd phase masks, where the symmetry is defined with respect to the pulse carrier frequency $\omega_0$. This symmetry plays an important role in coherent control experiments as pulse sequences with the same amplitude profiles but different temporal phase patterns can lead to strikingly different outcomes~\cite{Meshulach1998,Barmes2012}. 

\begin{figure*}
\includegraphics[width=\textwidth]{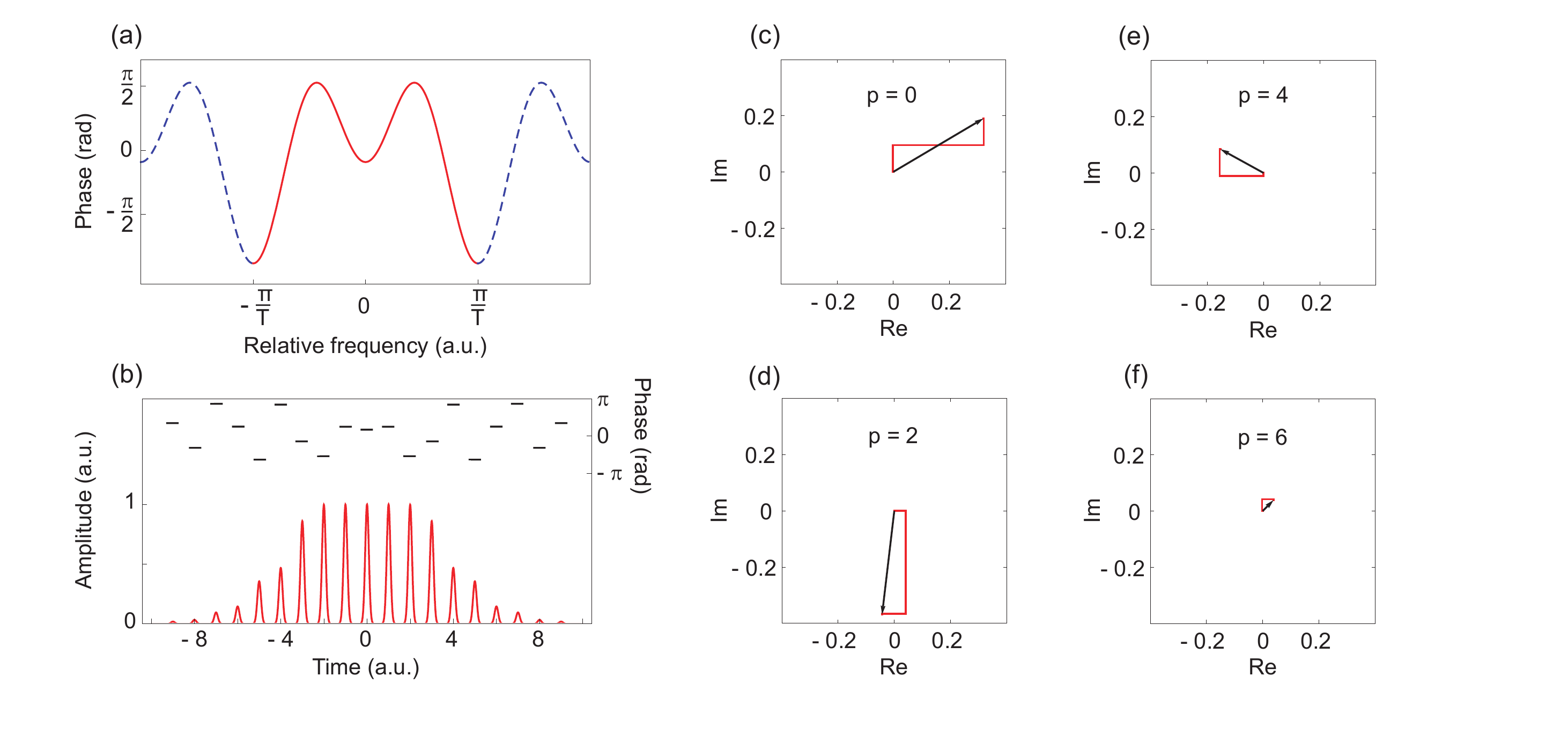}
\caption{Generation of pulse sequences by a periodic phase-only mask where a single period is an even function. (a) A periodic spectral phase mask as defined in Eq.~\ref{eq:periodic_even_mask} with $a=\{1.24,-1.53\}$ and $\Delta\omega=0$. (b) The amplitude and phase patterns generated by this phase mask. Vector representation of the individual amplitudes and phases contributing to the (c) 0$^{th}$ (d) 2$^{nd}$ (e) 4$^{th}$ and (f) 6$^{th}$ subpulses.}
\label{fig:2_coefficients}
\end{figure*}

In ultrafast optics~\cite{Trager2007} the (positive frequency) electric field of a short pulse with carrier frequency $\omega_0$ is written as $E^+_{in}(t)=\mathcal{E}^+_{in}(t) e^{i\omega_0 t}$. The envelope function $\mathcal{E}^+_{in}(t)$ encompasses both the amplitude and the nonlinear phase of the pulse and is generally a complex function. The generation of pulse sequences occurs when a periodic spectral phase mask of the form
\begin{equation}
\label{eq:periodic_even_mask}
\varphi(\omega)=\sum_{k=1}^{r} a_k \cos \left [kT(\omega-\omega_{ref})\right],
\end{equation}
with periodicity $\frac{2\pi}{T}$, is applied to a spectrum $E_{in}^+(\omega)$ (initially assumed to be Fourier-limited). The output field can be written as:
\begin{equation}
\begin{split}
E_{out}(\omega)&=E^{+}_{in}(\omega)e^{ i \sum_{k=1}^{r} a_k \cos \left [kT (\omega-\omega_{ref})\right]}\\
&=E^+_{in}(\omega)\prod_{k=1}^{r}e^{ i  a_k \cos \left [kT (\omega-\omega_{ref})\right]},
\end{split}
\label{eq:frequency_domain_field}
\end{equation}
where $\omega_{ref}$ determines the origin of the phase modulation. In order to find an expression for the temporal waveform we need to perform a Fourier transformation of Eq.~\ref{eq:frequency_domain_field}. Before this can be done the expression needs to be simplified. The first step is to use the Jacobi-Anger relation
\begin{equation}
\label{eq:Jacobi Anger sin}
e^{iA\cos(\theta)}=\sum_{n=-\infty}^{\infty}i^{n}J_{n}(A)e^{in\theta}.
\end{equation}
Inserting Eq.~\ref{eq:Jacobi Anger sin} into Eq.~\ref{eq:frequency_domain_field} leads to:
\begin{equation}
\label{eq:with Jacobi_Anger}
E^+_{out}(\omega)=E^+_{in}(\omega)\prod_{k=1}^{r}  \left[ \sum_{n=-\infty}^{\infty} i^n J_{n}(a_k) e^{i nkT (\omega-\omega_{ref})}\right].
\end{equation}
As a second step we rearrange the terms by writing the product of sums as a sum of products, which gives
\begin{equation}
\label{eq:exchange sum product}
\begin{split}
E^+_{out}(\omega)= \sum_{n_1,..., n_r=-\infty}^{\infty} \left[\prod_{k=1}^{r} i^{n_k} J_{n_k}(a_k)\right]\\
\times e^{ i T(\omega-\omega_{ref})\sum_{k=1}^{r} n_k k } E^+_{in}(\omega).
\end{split}
\end{equation}
Grouping these terms together allows us to perform the Fourier transformation to each term individually. The temporal profile of the shaped pulse now reads
\begin{equation}
\label{eq:time_domain_field}
\begin{split}
E^+_{out}(t)=e^{i\omega_0 t}\sum_{n_1,..., n_r=-\infty}^{\infty} &\left(\prod_{k=1}^{r} i^{n_k} J_{n_k}(a_k) \right )\\
&\left (e^{i \Delta \omega T\sum n_k k} \right )\\
&\mathcal{E}^+_{in}(t+T \sum n_k k),
\end{split}
\end{equation}
where $\Delta \omega=\omega_0-\omega_{ref}$ is the frequency detuning between the periodic phase mask and the laser carrier frequency. Each term in Eq.~\ref{eq:time_domain_field} corresponds to a set of the indices $\{n_1,...,n_r \}$, and represents a time-delayed scaled replica of the original pulse. It is characterized by the complex envelope function
\begin{equation}
\label{eq:complex_amplitude}
\begin{split}
\mathcal{E}^+_{out}(t,n_1,...,n_r)= &\left(\prod_{k=1}^{r} i^{n_k} J_{n_k}(a_k) \right )\\
&\left (e^{i \Delta\omega T\sum n_k k} \right )\\
&\mathcal{E}^+_{in}(t+ T \sum n_k k),
\end{split}
\end{equation}
which defines the amplitude and phase for each subpulse. Provided that  T is larger than the pulse duration of the unmodulated pulse, Eq.~\ref{eq:time_domain_field} represents a sequence of well separates subpulses where the p$^{th}$ subpulse is computed by summing over all terms for which $\sum n_k k=p$.

An example of a periodic spectral phase mask with parameters $a_k=\{1.24,-1.53\}$ and $\Delta\omega=0$ is shown in Fig.~\ref{fig:2_coefficients}a. The corresponding pulse sequence, as given by Eq.\ref{eq:time_domain_field}, is plotted in Fig.~\ref{fig:2_coefficients}b where the summation over the indices $n_k$ is truncated to include only the terms where $-10<n_k<10$. A  vector representation of the individual terms leading to the same subpulse are shown in Fig.~\ref{fig:2_coefficients}(c-f) for a number of selected subpulses. As clearly shown in the figure, only a small number of terms actually contributes significantly to the total amplitude. The amplitude pattern of the pulse sequence in Fig.~\ref{fig:2_coefficients}b is symmetric with respect to t=0. This is a general property of pulse sequences generated by a periodic phase mask where a single period is an even function. To understand this we note that Eq.~\ref{eq:time_domain_field} consists of pairs of terms with $\{n_k\}$ and $\{n_k'\}=(-1)\cdot \{n_k\}$ contributing to the p$^{th}$ and -p$^{th}$ subpulse, respectively. The amplitudes of these two terms are equal due to the relation $J_{-n}(x)=(-1)^{n}J_n(x)$. The same argument holds for the temporal phase pattern, provided that $\Delta \omega=0$. As can be seen from Eq.~\ref{eq:complex_amplitude} a nonzero $\Delta\omega$ adds a phase to each subpulse that is linear with respect to the pulse number $p=\sum n_k k$. While varying $\Delta\omega$ the vectors in Figs.~\ref{fig:2_coefficients}(c-f) rotate, leaving the amplitude of the subpulses unchanged. Naturally, after $\Delta\omega$ is changed by a full period the original phase pattern is recovered.


So far we considered periodic phase masks where a single period is an even function. We now turn to the case where the periodic phase mask is composed of odd functions
\begin{equation}
\label{eq:odd_periodic_function}
\varphi(\omega)=\sum_{k=1}^{r} a_k \sin \left [kT (\omega-\omega_{ref})\right].
\end{equation}
The derivation of the time domain waveform resulting from this spectral phase mask is similar to the one presented above, with one distinct difference. The Jacobi-Anger relation for odd sinusoidal functions is
\begin{equation}
e^{iA\sin(\theta)}=\sum_{n=-\infty}^{\infty}J_{n}(A)e^{in\theta}.
\end{equation}
Following the same steps as in Eq.~\ref{eq:with Jacobi_Anger} and Eq.~\ref{eq:exchange sum product} we find the time domain electric field to be
\begin{equation}
\label{eq:time_domain_field_odd_mask}
\begin{split}
E^+_{out}(t)=e^{i\omega_0 t}\sum_{n_1.... n_r=-\infty}^{\infty} &\left(\prod_{k=1}^{r}J_{n_k}(a_k) \right )\\
&\left (e^{ -i \Delta \omega T\sum n_k k} \right )\\
&\mathcal{E}^+_{in}(t+T \sum n_k k).
\end{split}
\end{equation}
As in the previous case, this leads to a sequence of subpulses where each subpulse is computed by summing over all complex envelope contributions
\begin{equation}
\label{eq:complex_amplitude_odd_mask}
\begin{split}
\mathcal{E}^+_{out}(t,n_1,...,n_r)= &\left(\prod_{k=1}^{r} J_{n_k}(a_k) \right )\\
&\left (e^{i \Delta\omega T\sum n_k k} \right )\\
&\mathcal{E}^+_{in}(t+T \sum n_k k).
\end{split}
\end{equation}
for which $\sum n_k k=p$.

\begin{figure}
\includegraphics[width=\columnwidth]{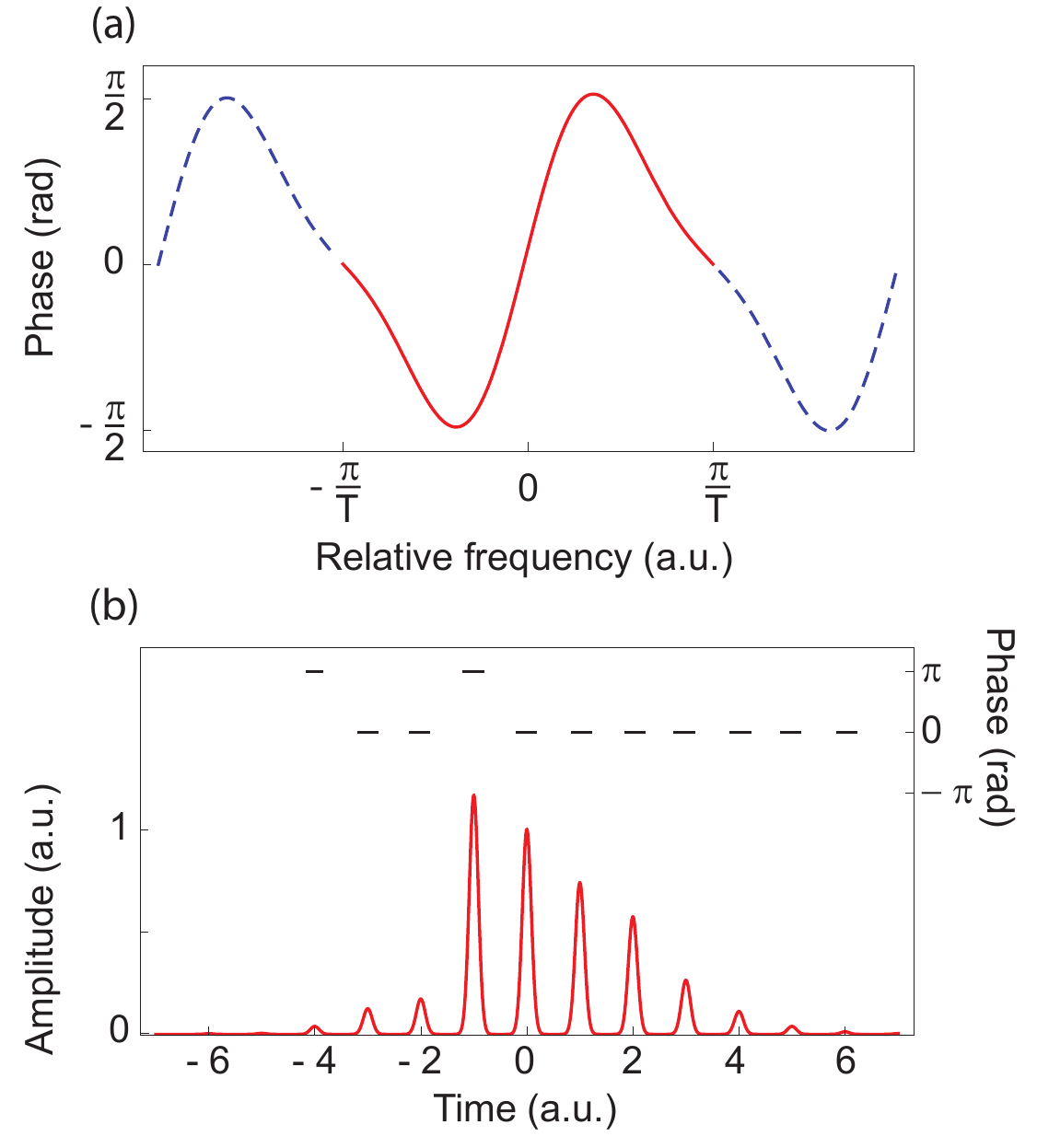}
\caption{Generation of asymmetric pulse sequences with an odd spectral phase mask. (a) A spectral phase mask as defined in Eq.~\ref{eq:odd_periodic_function}, with $a=\{1.4,0.4\}$ and $\Delta\omega=0$. (b) The resulting temporal amplitude and phase patterns}
\label{fig:odd_periodic_phase_mask}
\end{figure}
Equations~\ref{eq:time_domain_field} and~\ref{eq:time_domain_field_odd_mask} are almost identical. However, the pulse sequences generated by odd and even base functions exhibit markedly different properties. Most notable, the absence of the $i^{n_k}$ in Eq.~\ref{eq:complex_amplitude_odd_mask} means that the amplitudes of the p$^{th}$ and -p$^{th}$ subpulses are no longer necessarily equal. The possibility to generate asymmetric pulse sequences is illustrated in Fig.~\ref{fig:odd_periodic_phase_mask} for the Fourier coefficients $a_k$=\{1.4,0.4\} and $\Delta\omega=0$. In this example a linear increase between consecutive subpulses is clearly visible, but other sequences are possible by optimizing the values of the Fourier coefficients for a specific goal. An additional difference from the case of an even phase mask is that the complex envelope is a real function with temporal phase being only 0 or $\pi$. A nonzero $\Delta\omega$ has the same influence as explained above.


\begin{figure}
\includegraphics[width=\columnwidth]{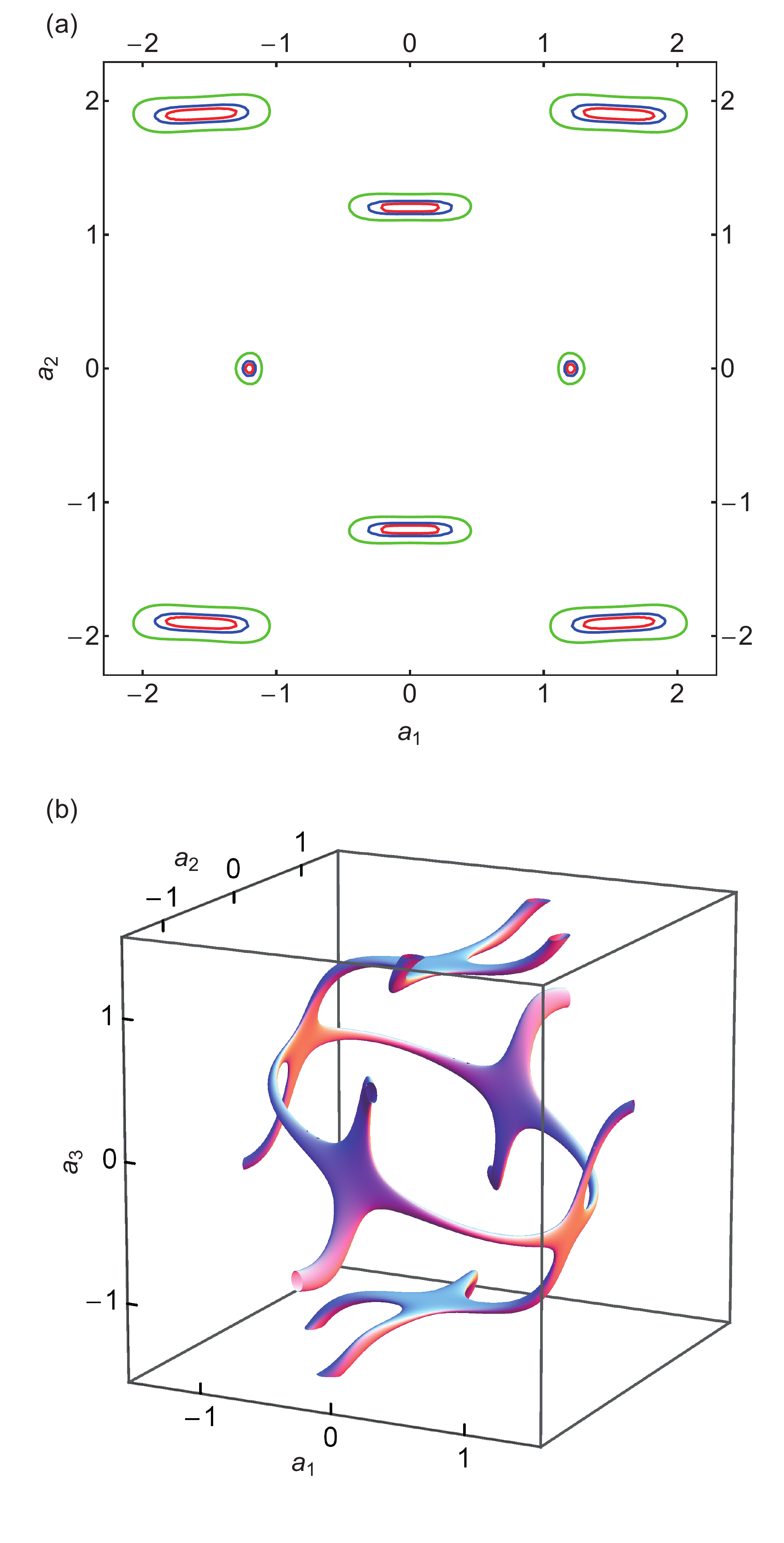}
\caption{Control landscape of dark pulses in nonresonant TPA. (a) In two dimensions the contours of excitation at 0.1$\%$ 0.25$\%$ and 1$\%$ percent are displayed in red blue and green, respectively. (b) A 3D contour plot of 0.1\% excitation level shows that all dark pulses are connected, displaying a unique topology of the control landscape.}
\label{fig:control_landscape}
\end{figure}

Parametrization of the spectral phase into a base of physically relevant functions has been employed for the investigation of the general properties of control landscapes~\cite{Rabitz2004}. We now illustrate how periodic spectral phase masks provide a viable parametrization for coherent control experiments by considering the process of nonresonant two-photon absorption. In this scheme a two-photon transition is excited by pairs of frequencies where no intermediate one-photon resonances lie within the laser bandwidth. In the weak-field regime, the transition amplitude is proportional to the integral
\begin{equation}
\label{eq:nonresonant_TPA}
\begin{split}
S=&\left |\int d\omega A(\omega)A(-\omega)\exp[i(\varphi(\omega)+\varphi(-\omega))]\right |^2,
\end{split}
\end{equation}
where $A(\omega)$ and $\varphi(\omega)$ are the spectral amplitude and phase of the driving field and the integration is performed relative to half the transition frequency. Meshulach \textit{et al}~\cite{Meshulach1998} have shown that the two-photon transition rate can be completely suppressed by a specific choice of pulse sequences. These so-called "dark pulses" are acquired when the modulation amplitude $A$ of a sinusoidal phase mask satisfies $J_0(2A)=0$ and the phase mask is an even function with respect to half of the transition frequency. In order to find general properties for dark pulses induced by arbitrary periodic spectral phase masks, we insert Eq.~\ref{eq:periodic_even_mask} into Eq.~\ref{eq:nonresonant_TPA} and follow the the same steps as in Eq.~\ref{eq:with Jacobi_Anger} and \ref{eq:exchange sum product}. The transition amplitude then reads
\begin{equation}
\begin{split}
S=&\left |\sum_{n_1,..,n_r}\prod_k \left ( i^{n_k}J_{n_k}(2 a_k)\right )\right.\\
&\left.\times\int d\omega A(\omega)A(-\omega)e^{-iT\omega \sum_k n_k k}\right|^2.
\end{split}
\label{eq:TPA_harmonic_phase}
\end{equation}
In the case of well separated subpulses the integrand in Eq.~\ref{eq:TPA_harmonic_phase} oscillates rapidly leading to negligible contribution to the excitation. Only terms for which $\sum n_k k=0$ are nonzero which means that the transition probability depends solely on the intensity of the central subpulse. The problem of finding dark pulses is therefore reduced to computing the amplitude of the central subpulse of Eq.~\ref{eq:time_domain_field}. Figure~\ref{fig:control_landscape}a displays a two-dimensional contour plot of 0.1$\%$, 0.25$\%$ and 1$\%$ excitation level (compared to the signal acquired by Fourier-limited pulses). Multiple solutions for zero transition amplitude (dark pulses) are observed, displaying local traps in the control landscape. However, it is expected that an unconstrained control landscape should exhibit a trap-free behavior~\cite{Roslund2009}. Indeed, allowing an additional control parameter shows that all dark pulse solutions are connected in a three dimensional landscape, displaying a trap-free landscape (Fig.~\ref{fig:control_landscape}b). The topology of this control landscape presented here is very different to the one acquired by polynomials~\cite{Roslund2006,Roslund2009}. It can therefore serve as a complementary test for studies of the general properties of control experiments.

In conclusion we have introduced a new class of spectral phase masks for coherent control experiments, namely, arbitrary periodic phase functions. A derivation of the temporal waveform, relating the temporal amplitudes and phases of the subpulses to the Fourier coefficients, provides insight to the control mechanism of the quantum system. This method can be readily employed in many existing experiments what were thus far limited to a single sinusoidal phase mask.


\bibliography{biblio}

\end{document}